\documentclass[pss]{wiley2sp}
\usepackage{comment}
\usepackage{color}
\usepackage{bm}
\usepackage{w-greek}

\begin{document}

\title{Spin textures of strongly correlated spin Hall quantum dots}
\titlerunning{Spin textures of strongly correlated quantum spin Hall dots}

\author{
  Giacomo Dolcetto\textsuperscript{\textsf{\bfseries 1,2,3}},
  Niccol\`o Traverso Ziani\textsuperscript{\textsf{\bfseries 1,2}},
  Matteo Biggio\textsuperscript{\textsf{\bfseries 4}},
  Fabio Cavaliere\textsuperscript{\textsf{\Ast,\bfseries 1,2}},
  Maura Sassetti\textsuperscript{\textsf{\bfseries 1,2}}}

\authorrunning{G. Dolcetto {\em et al}.}

\mail{e-mail
  \textsf{fabio.cavaliere@gmail.com}, Phone:
  +39-010-3536248, Fax: +39-010-314218}

\institute{
  \textsuperscript{1}\,Dipartimento di Fisica, Universit\` a di Genova, Via Dodecaneso 33, 16146, Genova, Italy\\
  \textsuperscript{2}\,CNR-SPIN, Via Dodecaneso 33, 16146, Genova, Italy\\
  \textsuperscript{3}\,INFN, Via Dodecaneso 33, 16146, Genova, Italy\\
  \textsuperscript{4}\,Dipartimento di Ingegneria Navale, Elettrica, Elettronica e delle Telecomunicazioni, Universit\` a di Genova, Via Opera Pia 11a, 16145, Genova, Italy}
 
\received{XXXX, revised XXXX, accepted XXXX} 
\published{XXXX} 


\abstract{{We study the spin ordering of a quantum dot defined via magnetic barriers in an interacting quantum spin Hall edge. The spin-resolved density-density correlation functions are computed. We show that strong electron interactions induce a ground state with a highly correlated spin pattern. The crossover from the liquid-type correlations at weak interactions to the ground state spin texture found at strong interactions parallels the formation of a one-dimensional Wigner molecule in an ordinary strongly interacting quantum dot.}}
\titlefigure[height=3.5cm,keepaspectratio]{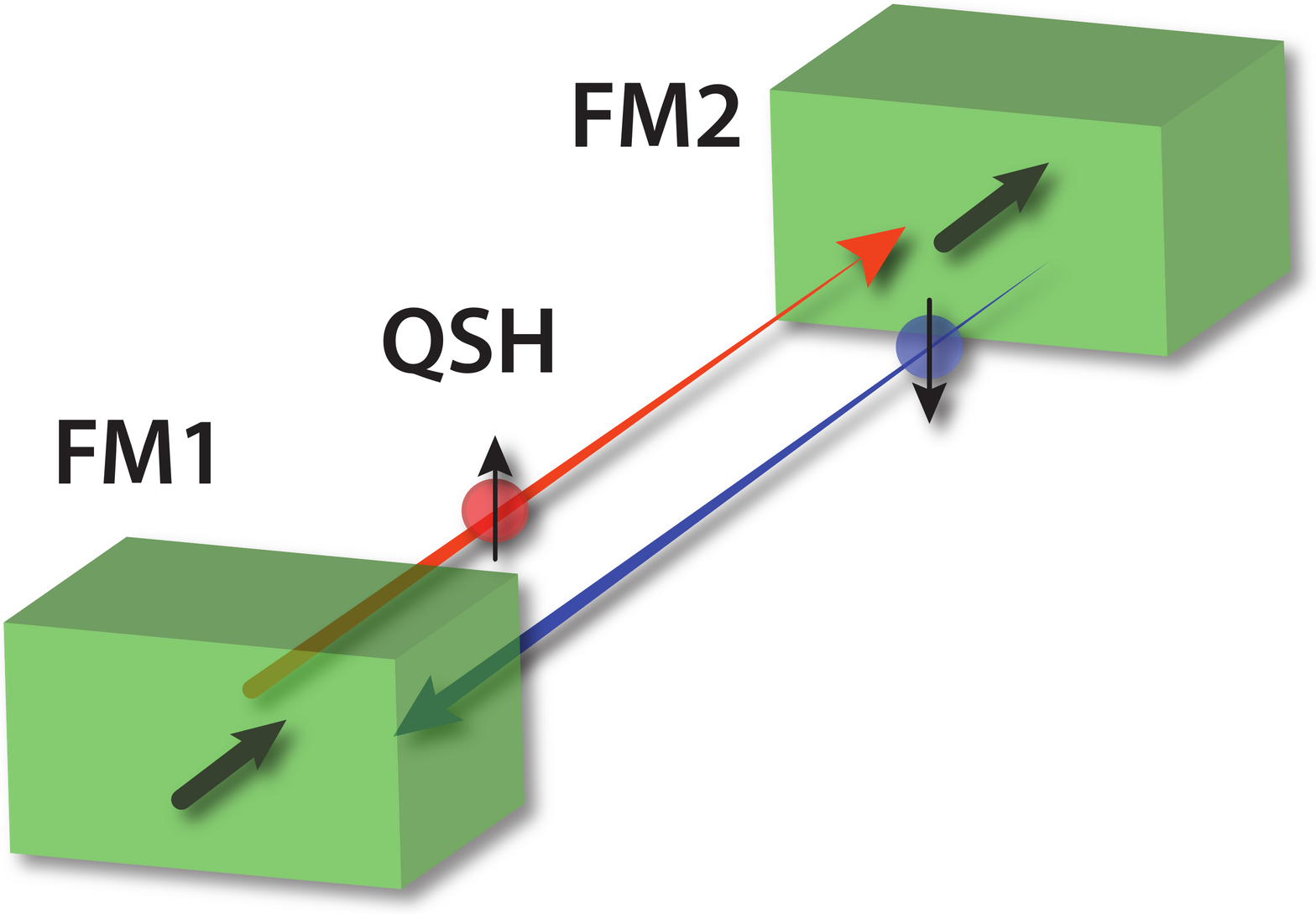}
\titlefigurecaption{A quantum dot, delimited by two magnetic barriers, embedded into a quantum spin Hall edge.}
\maketitle  
A topological insulator (TI) \cite{qizhang,hasankane} is a material that behaves like an ordinary band insulator in the bulk,
but has gapless states near its boundaries \cite{rothscience,buttiker}. In the quantum spin Hall (QSH) phase \cite{kanemele,bernevigscience}, realized in two-dimensional
TIs, edge states are characterized by spin-momentum locking \cite{rothscience,brune,wubernevigzhang} and, as long as time-reverval (TR) symmetry is preserved, are topologically robust against single-particle backscattering \cite{wubernevigzhang,budich,Schmidt}.
The experimental evidence of ballistic helical transport \cite{rothscience,brune,konig} paved the way for studying transport properties in different geometries, ranging form the standard quantum point contact geometry \cite{strom} to more sophisticated interferometric setups \cite{interferometric}. Other tools such as the magnetic susceptibility~\cite{loss}, or transport properties involving local probes such as a STM tip~\cite{das} have also been proposed. Furthermore, two-dimensional TIs are considered as promising building blocks for spintronic applications \cite{spintronics}, for the search of Majorana fermions \cite{majorana} and in the context of quantum information \cite{computation}.\\
\noindent In order to study the topologically non-trivial nature of the edge states, various authors have suggested the possibility of realizing quantum dots as finite-length portions of QSH edges \cite{qinature,timm,dolcetto2013coulomb}.
{A possible way to confine the edge states involves electrical control of topological phase transitions in double HgTe/CdTe ~\cite{doubleQW} or InAs/GaSb ~\cite{singleQW} quantum wells. In addition,} one can have spin-dependent scattering induced by magnetic materials which opens a gap, thus leading to the formation of a dot \cite{timm,dolcetto2013coulomb}. In particular, the ground-state properties of a QSH quantum dot realized by two opaque ferromagnetic barriers have recently been the subject of intense studies. The time-reversal symmetry breaking due to magnetic impurities~\cite{add1,add2,add3,add4} leads to interesting and novel effects. For instance, Qi \emph{et al.} \cite{qinature} showed that a magnetic domain wall can induce a localized fractional charge on a single QSH edge, and suggested a Coulomb blockade setup to detect it. Furthermore, an oscillating pattern arises in the in-plane spin density as a consequence of helicity combined with the presence of magnetic barriers confining the dot~\cite{dolcetto2013coulomb,Meng12}, strongly enhanced in the presence of electron interactions. {The competition between Friedel oscillations and the formation of a Kondo cloud at low temperature in a QSH state has also been recently considered~\cite{kondo}.}\\
\noindent {The oscillating spin pattern mentioned above parallels the charge density oscillations in an ordinary quantum dot.
In the latter case, the charge density profile exhibits oscillations with $4k_F$ periodicity~\cite{wig1234,wig5,wig6,wig7,wig8}. However, just like the oscillating charge density profile is not by itself an indication of true Wigner crystallization~\cite{vignale,polinirontani}, the appearence of spin-density oscillations in QSH dots cannot be interpreted as a clear indication of a ground state with short-range spin ordering.
To investigate the formation of a true correlated spin structure of the ground state, one needs to study spin-resolved density-density correlation functions. In view of the strong interest of such novel systems for innovative applications including spintronic, a thorough investigation of the spin correlations is of paramount importance both at the level of fundamental research and as the possible trigger for devising novel QSH-based devices.}\\
\noindent In this Letter we elucidate the role of electron interaction in the formation of a spin ordered state. We confirm, by analyzing the quantum dot pair correlation functions, that electron interactions lead to the formation of a strongly correlated spin state. We demonstrate that by increasing the interactions, a crossover from an uncorrelated spin state towards a state displaying a well defined ground state spin texture is found.\\

\begin{figure}[htbp]
\begin{center}
\includegraphics[width=8cm,keepaspectratio]{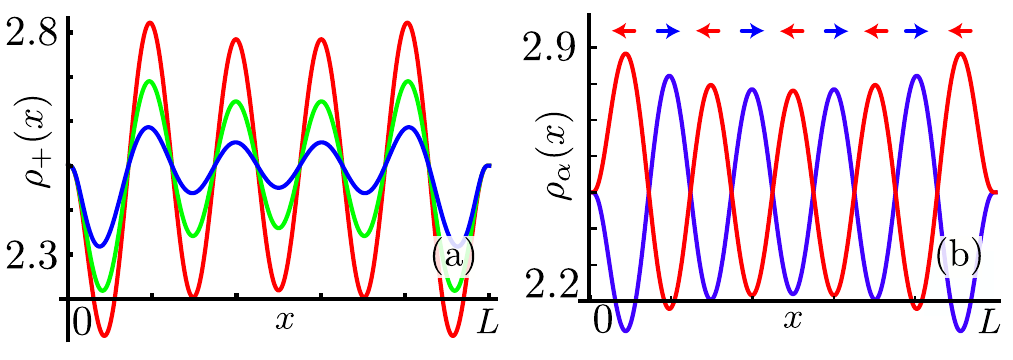}
\caption{(a) Plot of $\rho_{+}(x)$ (units $1/L$) as a function of $x$ for $K=0.25$ (red), $K=0.55$ (green), $K=1$ (blue) and (b) plot of $\rho_{\alpha}(x)$ (units $1/L$) as a function of $x$ for $K=0.25$ for $\alpha=+$ (blue) and $\alpha=-$ (red). Both panels are for $5$ electrons in the dot, $T=0$ and $\alpha/L=1/16$.}
\label{figura1}
\end{center}
\end{figure}

\begin{figure}[htbp]
\begin{center}
\includegraphics[width=8cm,keepaspectratio]{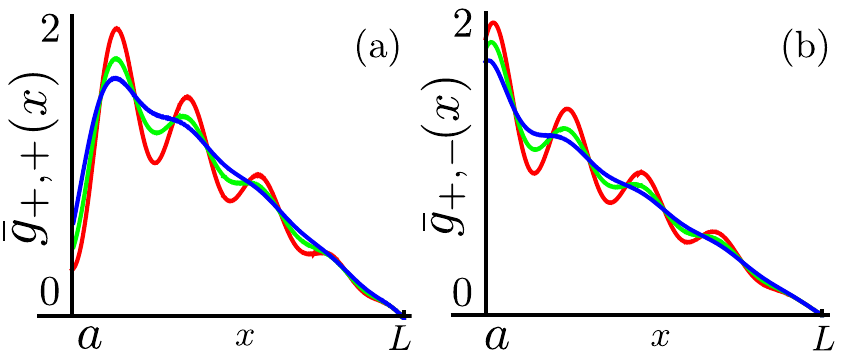}
\caption{Plot of (a) $\bar{g}_{+,+}(x)$ and (b) $\bar{g}_{+,-}(x)$ (units $1/L$) as a function of $x$ for $K=0.25$ (red), $K=0.55$ (green), $K=1$ (blue) for $5$ electrons in the dot. Here, $T=0$ and $\alpha/L=1/16$.}
\label{figura2}
\end{center}
\end{figure}

\begin{figure}[htbp]
\begin{center}
\includegraphics[width=7cm,keepaspectratio]{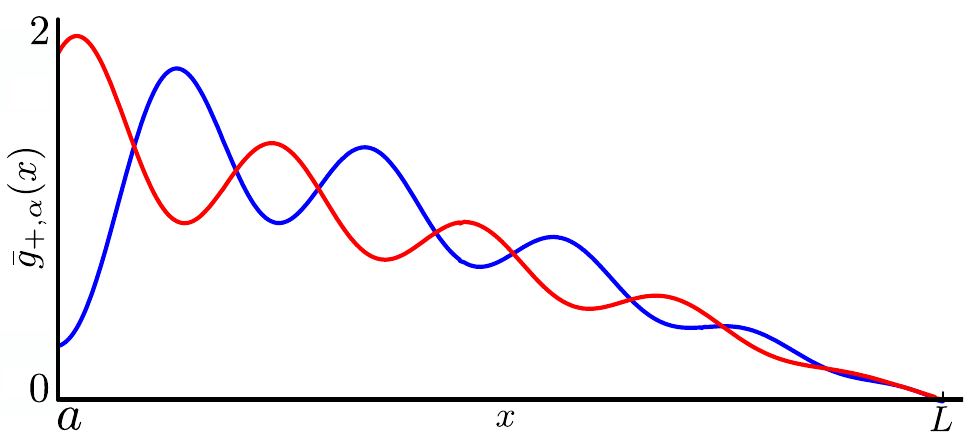}
\caption{Plot of $\bar{g}_{+,\alpha}(x)$ (units $1/L$) for $\alpha=+$ (blue) and $\alpha=-$ (red) as a function of $x$ for $K=0.25$ and $5$ electrons in the dot. Here, $T=0$ and $\alpha/L=1/16$.}
\label{figura4}
\end{center}
\end{figure}

\noindent The system consists of a quantum dot built in a QSH helical edge state characterized by the free Hamiltonian ($\hbar=1$)
\begin{equation}\label{freefirst}
	H_{0} = -iv_{F}\sigma_{z}\partial_{x},
\end{equation}
where $v_{F}$ is the Fermi velocity, $\vec{\sigma}=\left(\sigma_x,\sigma_y,\sigma_z\right)$ are the Pauli matrices, and $x$ is the coordinate along the edge. A propagating edge state is described by the single-particle spinorial wavefunction $\Psi(x)= \left(\Psi_{R\uparrow}(x)\,\,\, \Psi_{L\downarrow}(x)\right)^{T}$, where $\Psi_{R\uparrow(L\downarrow)}(x)$ represents a right (left) moving electron with spin-up (-down) along the $\hat{z}$ direction. Two thin localized ferromagnetic barriers, with parallel magnetizations pointing in the $\hat{x}$ direction, are modeled by two delta functions placed at $x =0$ and $x=L$ with Hamiltonian
\begin{equation}\label{barriers}
	H_{FM} = -M\left[\delta(x)+\delta(x-L)\right]\sigma_{x},
\end{equation}
where $M$ is the magnetization strength. In the limit $M/v_{F}\to\infty$, the Hamiltonian in Eq.~(\ref{barriers}) decouples the wavefunctions defined in $[0,L]$ from the ones of the outer region, inducing peculiar boundary conditions~\cite{timm,dolcetto2013coulomb}
\begin{equation}\label{BCs}
\Psi_{L\downarrow}(0)=-i\Psi_{R\uparrow}(0), \ \ \ \
\Psi_{L\downarrow}(L)=i\Psi_{R\uparrow}(L).
\end{equation}
The solution of the Schr\"{o}dinger equation $H_{0}\Psi_{q} = E_{q}\Psi_{q}$ in $[0,L]$, togheter with the conditions in Eq.~(\ref{BCs}), gives the energy eigenvalues $E_{q} = v_{F}q$, where $q = \frac{\pi}{L}\left( n-\frac{1}{2} \right)$ are the quantized momenta, $n$ an integer, and $\Psi_{q}(x)$ are the corresponding eigenvectors ${{\Psi}_{q}(x) = \left ( e^{iqx} -ie^{-iqx}\right )^T/\sqrt{2}.}$
The electronic field operator $\hat{\Psi}(x)$ can thus be expanded on the basis $\left \{\Psi_{q}(x)\right \}$ as $\hat{\Psi}(x)=\sum_{q}\Psi_q(x)\hat{c}_q$, the operator $\hat{c}_q$ destroying an electron with energy $E_q$.
In particular, the two-component spinorial field operator (projected along $\hat{z}$) can be written as \cite{dolcetto2013coulomb}
\begin{equation}
{	\hat{\Psi}(x) = {\Psi_{\uparrow}(x) \choose \Psi_\downarrow(x)} = {e^{ik_{F}x}\hat{\psi}(-x)\choose -ie^{-ik_{F}x}\hat{\psi}(x)}}
\end{equation}
{where $\hat{\Psi}_{\sigma}$ are the up ($\sigma=\uparrow$) and down ($\sigma=\downarrow$) components along the $\hat{z}$ axis.}
The field $\hat{\psi}(x)$ satisfies antiperiodic boundary conditions over a segment of length $2L$, $\hat{\psi}(L) = -\hat{\psi}(-L)$, and can be bosonized as \cite{vondelft}
\begin{equation}
	\hat{\psi}(x) = \frac{{\mathcal{F}}}{\sqrt{2\pi a}} e^{-i\pi\frac{x}{L} \left( \Delta\hat{N} - \frac{1}{2} \right)}e^{-i\hat{\phi}(x)},
\end{equation}
where ${\mathcal{F}}$ is a Klein factor, $\Delta\hat{N} = \hat{N}-N_{0}$ is the excess number of particles with respect to $N_{0}$, and $\hat{\phi}(x) = - \sum_{k> 0} \sqrt{\frac{\pi}{kL}} e^{-\frac{\alpha}{2}k} \left[ e^{-ikx}\hat{b}_{k} + \mbox{h.c.}\right]$.
Here, $k=\frac{\pi n}{L}$, $\hat{b}_{k}$ is a bosonic operator, annihilating a collective density excitation with energy $v_F k$, and $\alpha = L/(\pi N)$ is a short distance cutoff.
We will consider the model Hamiltonian $\hat{H} = \hat{H}_{0} + \hat{H}_{int}$ for the edge states, with $\hat{H}_{0}$ expressed as
\begin{equation}
\hat{H}_{0} = -iv_{F}\int_{0}^{L}dx :\hat{\Psi}^{\dagger}(x) \sigma_{z} \partial_{x}\hat {\Psi}(x):
\end{equation}
in a form equivalent to Eq.~\ref{freefirst} and electron interactions modeled by
\begin{equation}
\hat{H}_{int} = \frac{g}{2} \int_{0}^{L} dx \left[ \hat{\Psi}^{\dagger}(x)\hat{\Psi}(x) - \frac{N_{0}}{L}\right]^{2}.
\end{equation}
By means of a Bogoliubov transformation the Hamiltonian $H$ can be diagonalized to
\begin{equation}\label{diagonalized}
\hat{H} = \epsilon \sum_{k>0}m(k)\hat{a}^{\dagger}_{k}\hat{a}_{k} + \frac{E_{N}}{2} \left( \Delta\hat{N}\right)^{2}.
\end{equation}
In Eq.~\ref{diagonalized}, $\hat{a}_k$ are bosonic operators, $m(k)=kL/\pi$ is an integer number, $\epsilon = \pi v_{F}/(KL)$ and $E_{N} = \pi v_{F}/(K^{2}L)$ are respectively the plasmonic and charging units of energy, and $K = \left[1+g/(\pi v_{F})\right]^{-1/2}$ is the Luttinger parameter~\cite{note}.\\
{As a consequence of both the spin-momentum locking of helical liquids and the presence of magnetic barriers, the average $z$-component of the spin operator $\langle\hat{s}_z(x)\rangle = \langle\hat{\Psi}^{\dagger}(x)\sigma_z\hat{\Psi}(x)\rangle/2 = \langle\hat{\rho}_{\uparrow}(x)-\hat{\rho}_{\downarrow}(x)\rangle/2=0$, with $\hat{\rho}_{\sigma}(x)=\hat{\Psi}^{\dagger}_{\sigma}(x)\hat{\Psi}_{\sigma}(x)$, where $\langle\ldots\rangle$ denotes the thermal average at zero temperature~\cite{Physscr}.
On the other hand, the \emph{in-plane} components of the average spin $\langle \hat{s}_{x,y}(x)\rangle=\langle\hat{\Psi}^{\dagger}(x)\sigma_{x,y}\hat{\Psi}(x)\rangle/2$ are non-vanishing~\cite{dolcetto2013coulomb,add2,Meng12}.
\begin{figure}[htbp]
\begin{center}
\includegraphics[width=6cm,keepaspectratio]{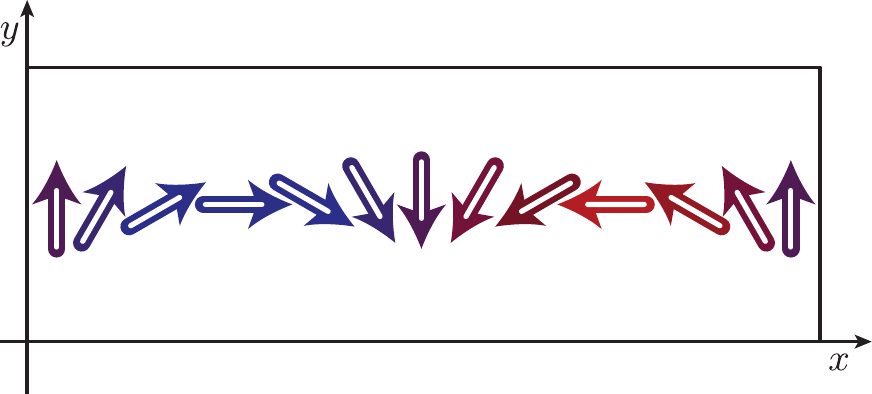}
\caption{Scheme of the in-plane average spin operator, which shows oscillations in its direction.}
\label{schema}
\end{center}
\end{figure}
In an ordinary spinfull Luttinger liquid with ferromagnetic barriers, the in-plane magnetization is generically dominated by long-wave, non-oscillating terms~\cite{Meng12}. However, such terms are absent in QSH systems due to helicity, and only oscillating terms like $\hat{\Psi}^{\dagger}_{\uparrow}(x)\hat{\Psi}_{\downarrow}(x)=e^{-i2k_Fx}\hat{\psi}^{\dagger}(x)\hat{\psi}(-x)$ contribute \cite{Meng12}. The presence of magnetic barriers, which break translational invariance and couple spin up and spin down components (see Eq.~(\ref{BCs})), then allow to have non-vanishing average values of the in-plane components of the spin operator, resulting in a peculiar average spin density oscillating pattern  schematically shown in Fig.~\ref{schema}\\
For this reason, we will focus on the $x$-projected spin-resolved electron densities $\hat{\rho}_{\alpha}=\hat{\Psi}^{\dagger}_{\alpha}\hat{\Psi}_{\alpha}$, where $\alpha=+$ ($-$) corresponds to spin up (down) along the $\hat{x}$ direction.
The relations between spin states in the $\hat{x}$ and $\hat{z}$ directions $\hat{\Psi}_{\alpha}(x)=\left[\hat{\Psi}_{\uparrow}(x)+\alpha\hat{\Psi}_{\downarrow}(x)\right]/\sqrt{2}$, allow to rewrite $\hat{\rho}_{\alpha}(x)=\hat{\rho}(x)/2+\alpha \hat{s}_x(x)$ with $\hat{\rho}(x)=\sum_{\alpha=\pm}\hat{\rho}_{\alpha}$ the total electron charge density.\\
\noindent Despite the average $\langle\hat{\rho}(x)\rangle = \left \langle\frac{\hat{N}}{L}+\frac{1}{2\pi}\partial_x\left [\hat{\phi}(x)+\hat{\phi}(-x)\right ]\right\rangle=N/L$ is constant, the average spin resolved electron densitiy shows an oscillating behavior with wavevector $2k_{F}$ due to~\cite{dolcetto2013coulomb}
$\langle\hat{s}_{x}(x)\rangle=\left\langle-ie^{2ik_{F}x}\hat{\psi}^{\dagger}(-x)\hat{\psi}(x)+\mathrm{H.c.}\right\rangle/2$.} The average $\rho_{\alpha}(x)=\langle \hat{\rho}_{\alpha}(x)\rangle$ then reads
\begin{equation}\rho_{\alpha}(x)=\frac{N}{2L}+\alpha s_x(x)
.\end{equation}
The evolution of the oscillations are shown in Fig.~\ref{figura1}(a): they occur even in the non-interacting case, become strongly enhanced at strong interactions ($K\to 0$), and are characterized by a typical wavelength $\delta\sim 2k_F\sim L/N$. {Figure~\ref{figura1}(b) compares $\rho_{+}(x)$ and $\rho_{-}(x)$, showing that their peaks are intercalated, with their sum constant.}
In order to understand if such spin oscillations signal the presence of ground state with true short-range spin correlations, we analyze the correlation function (CF) $\bar{g}_{\alpha,\alpha^{\prime}}(x)$ expressing the probability density of finding two electrons with spin projections $\alpha$ and $\alpha^{\prime}$ along the $\hat{x}$ axis at a relative distance $x$, given by~\cite{polinirontani}
\begin{equation}\label{CF}
	\bar{g}_{\alpha,\alpha^{\prime}}(x) = \int_{-\infty}^{\infty} dx^{\prime} g_{\alpha,\alpha^{\prime}}\left(x^{\prime}+\frac{x}{2},x^{\prime}-\frac{x}{2}\right)
\end{equation}
where~\cite{nota}
\begin{equation}\label{correlator}
	g_{\alpha,\alpha^{\prime}}(x,y) = \frac{\left\langle \hat{\Psi}_{\alpha}^{\dagger}(x)\hat{\Psi}_{\alpha^{\prime}}^{\dagger}(y)\hat{\Psi}_{\alpha^{\prime}}(y)\hat{\Psi}_{\alpha}(x) \right\rangle_{N}}{N(N-\delta_{\sigma,\sigma^{\prime}})}\, .
\end{equation}
For clarity, we restrict the analysis to the zero temperature case, in which the expectation values are taken with respect to the ground state. We explicitly checked that no qualitative changes occur up to temperatures $k_{B}T\sim \epsilon$, when thermal excitation becomes large enough to trigger the collective excitations of the system. Equation~(\ref{correlator}) can be calculated analytically using standard bosonization techniques, while the integration in Eq.~(\ref{CF}) has been performed numerically. 
{If a strongly correlated spin state arises, we expect to observe marked peaks in the CFs: in this case, we could effectively interpret the oscillations diplayed by the spin-resolved electron density as a genuine tendency towards spin ordering~\cite{polinirontani}.}\\
Figure~\ref{figura2}(a) shows $\bar{g}_{+,+}(x)$ as a function of $x$ for different values of the interaction strength. Notice that here, and in the following, we investigate correlations down to the shortest length scale $a$, the spatial cutoff of the theory. As a general feature, the parallel-spin CF displays a short-range suppression induced by the Pauli exclusion principle. For a noninteracting system (blue curve), an essentially featureless decay of $\bar{g}_{+,+}(x)$ is observed, reminiscent of an uncorrelated liquid-like state. On the other hand, as interactions increase strong oscillations develop, revealing a correlated spin state. The number of peaks is $N-1$, with average spacing given by $\delta$, the average wavelength of the density oscillations observed in Fig.~\ref{figura1}.
{At strong interactions, the oscillations displayed by the spin-resolved electron density really correspond to the arising of a ground state spin texture.}
The CF for anti-parallel spin orientations, shown in Fig.~\ref{figura2}(b), is qualitatively similar, but with the absence of the Pauli-induced short distance suppression.\\
Figure~\ref{figura4} shows a combined plot of the parallel- and antiparallel-spins CFs. {Crucially, the position of the peaks of $\bar{g}_{+,-}$ is shifted with respect to $\bar{g}_{+,+}$: an intercalated pattern of oscillations emerges, revealing the presence of a ground state spin texture at strong interactions.}
{This confirms the physical picture suggested by the spin-resolved electron densities $\rho_{\alpha}(x)$ in Figure~\ref{figura1}(b), and schematically represented in Fig.~\ref{schema}, with the probability of finding a nearest neighbour with the same spin being maximal at a distance twice as large with respect to that of finding a nearest-neighbour with opposite spin.} We have checked that the features described are independent from the number of electrons in the dot.\\
\noindent In conclusion, we have investigated the spin-spin correlations in a quantum dot built in a quantum spin Hall system. We have shown that for strong electron interactions the spin oscillations, already reported in Ref.~\cite{dolcetto2013coulomb}, are indeed the signature of a true magnetic order with strongly correlated spin state. The underlying spin structure closely resembles a state which could be described as a "spin molecule", in analogy with the formation of a Wigner molecule in a conventional quantum dot.
\begin{acknowledgement}
We thank D. Ferraro, A. Braggio and T. L. Schmidt for useful discussions. The support of EU-FP7 via Grant
No. ITN-2008-234970 NANOCTM is acknowledged.
\end{acknowledgement}


\begin{thebibliography}{[1]}
\bibitem{note} We consider here non-strictly-zero range interactions among electrons.
\bibitem{nota}
Note that, since the elctronic operator is defined in $[0,L]$, the function $g_{\sigma,\sigma '}(x,y)=0$ if $x,y<0$ or $x,y>L$.
\end{thebibliography}
\end{document}